\documentclass[12pt]{iopart}
\usepackage[dvips]{graphicx}
\usepackage{axodraw}
\usepackage{amssymb}
\newcommand{\be}{\begin{equation}}
\newcommand{\ee}{\end{equation}}
\newcommand{\bea}{\begin{eqnarray}}
\newcommand{\eea}{\end{eqnarray}}

\newcommand{\ket}[1]{\vert{#1}\rangle}

\def\MCN{{\mathcal N}}

\makeatletter
\renewcommand\section{\@startsection {section}{1}{\z@}%
                                   {-5.5ex \@plus -1ex \@minus -.2ex}
                                   {2.3ex \@plus.2ex}%
                                   {\normalfont\large\bfseries}}
\renewcommand\subsection{\@startsection{subsection}{2}{\z@}%
                                     {-3.25ex\@plus -1ex \@minus -.2ex}%
                                     {1.5ex \@plus .2ex}%
                                     {\normalfont\normalsize\bfseries}}
\renewcommand\thesection {\@arabic\c@section}
\renewcommand\thesubsection   {\thesection.\@arabic\c@subsection}
\renewcommand{\@seccntformat}[1]{%
\csname the#1\endcsname.\hspace{1.0em}}
\makeatother

\begin{document}

\begin{flushright}
HIP-2007-34/TH\\
\end{flushright}

\title{On the origin of thermal string gas}

\author{Kari Enqvist, Niko Jokela, Esko Keski-Vakkuri and Lotta Mether}

\address{Department of Physical Sciences, University of Helsinki \\ and
Helsinki Institute of Physics, \\ P. O. Box 64, FIN-00014 University of
Helsinki, Finland.}

\begin{abstract}
We investigate decaying D-branes as the origin of the thermal string gas
of string gas cosmology.  We consider initial configurations of low-dimensional
branes and argue that they can time evolve to thermal string
gas. We find that there is a range in the weak string coupling and
fast brane decay time regimes, where the initial configuration could
drive the evolution of the dilaton to values, where exactly three
spacelike directions grow large.
\end{abstract}

\eads{\mailto{kari.enqvist@helsinki.fi},
\mailto{niko.jokela@helsinki.fi},
\mailto{esko.keski-vakkuri@helsinki.fi},
\mailto{lotta.mether@helsinki.fi}}

\maketitle

\section{Introduction}

The origin of the primordial perturbations is one of the key
problems in cosmology. A scale invariant spectrum could indicate an
early de Sitter universe with a large cosmological constant, but the
latest WMAP data \cite{Spergel:2006hy} with a spectral index of
$n_s=0.958 \pm 0.016$ very much favors a dynamical origin. There
should be an order parameter, a field, whose value changes during
inflation and thus gives rise to a slight deviation from the exact
scale invariance. This is of course what slow-roll inflation models
\cite{linde} are famous for, but there could be alternatives to
scalar field inflation. In particular, string theoretical effects
could manifest themselves in the early universe, making the
primordial perturbation a testing ground for fundamental physics.
Attempts along these lines include pre-Big Bang cosmology
\cite{preBB} and various scenarios that involve dynamics of branes
\cite{Dbranesetc}, as well as string gas cosmology \cite{sgcreview},
which assumes that the early universe consisted of a collection of
strings in thermal equilibrium. String gas cosmology originates in
an early proposal of Brandenberger and Vafa
\cite{Brandenberger:1988aj} for a natural explanation of the
dimensionality of spacetime. Since then, the scenario has been
developed in various directions and the framework has been applied
to address other cosmological issues such as dark matter and, most
recently, structure formation \cite{stringgas}.

String gas cosmology focuses in particular on the effect of the
string winding degrees of freedom and the implications of T-duality.
The basic postulate of the model is that the universe starts out as
a small compact space, usually taken to be a string size torus for
simplicity, which is filled with a thermal gas of closed strings at
a temperature near the Hagedorn temperature $T_H$, which is believed
to be the maximal temperature for a perturbative string ensemble.
Already a simple qualitative consideration of the string gas
scenario has some very fundamental implications on the picture of
the early universe. The degrees of freedom in the string gas consist
of the three types of closed string modes: momentum modes
(describing the center of mass motion of strings), winding modes
(expressing the number of times a string is wound around a given
torus one cycle), and oscillatory modes. The energy of momentum
modes is quantized in units of the inverse torus radii, while that
of winding modes is directly proportional to the radii and the
oscillatory modes have radius independent energies. Thus, the string
spectrum is invariant under T-duality, \emph{i.e.}, the inversion of
the torus radii, along with the interchange of the corresponding
momentum and winding quantum numbers. This symmetry implies the
existence of an effective minimal length scale, which in turn
suggests a resolution of both the spatial and temperature
singularities of the early universe.

But what could be the origin of the thermal string gas? Let us break
the question of the initial conditions into two subquestions: (i)
the origin of the spacetime itself, and (ii) the origin of the
matter or energy in the spacetime. The first one is the question of
how to resolve the initial spacelike Big Bang singularity. The most
famous scenario is the no-boundary proposal \cite{Hartle:1983ai},
but there have been other more recent approaches. We only mention
the idea of disappearance of space by closed string tachyon
condensation, the time reversed process interpreted as emergence of
space from nothing \cite{tachyoncosmo} (see also
\cite{moretachyon}). This scenario has recently been connected with
string gas cosmology in \cite{Brandenberger:2007xu}. In this paper we address 
the second question. How did the hot string gas come
to exist -- what creates the flux of energy from the initial
singularity?\footnote{Note that the second paper in \cite{Brandenberger:2007xu} also addresses this question and presents an argument for creation of a thermal distribution. However, the approach in \cite{Brandenberger:2007xu} is very different from ours. For an oscillatory scenario, see \cite{Biswas:2006bs}; for one
 involving branes but which differs from the setup in the present paper, see \cite{Chowdhury:2006pk}.}

In a recently proposed scenario, energy in a spacetime can originate
from a decaying brane, for which an initial condition can be
prepared at the origin of time \cite{Kawai:2005jx}. Loosely
speaking, the spacetime comes to existence with a great amount of
stored energy which gets released immediately. It is also
conceptually interesting that brane decay is an example of a process
which can be replaced by a microscopic statistical model, where time
is absent and thus ``time evolution'' can be modeled as an emergent
concept \cite{Balasubramanian:2006sg}.

In the present paper we study specifically if a thermal gas of
closed strings could arise as a consequence of the decay of unstable
D-branes. This then provides a specific initial condition scenario
for the thermal string gas in string gas cosmology at a {\em finite}
origin of time (rather than introducing the thermal gas {\em by
fiat}). One virtue of the scenario is that there is freedom in
choosing the details of the initial condition, and this is reflected
in the subsequent development. For example, a space filling brane
appears to produce string-brane gas \cite{Alexander:2000xv}, while a
configuration of lower dimensional branes can produce pure string
gas. The details of the thermalization leave room for more detailed
analysis and possible interesting effects. We focus on the time
evolution of the dilaton and the string coupling. We find that the
evolution is internally consistent and can lead to favorable values
of the dilaton for three dimensions to grow large.

There then exist two complementary scenarios: emergence of spacetime
by closed string tachyon condensation, and emergence of matter in
the spacetime by open string tachyon condensation (brane decay).
Future work will show if and how the two can be fused together for a
complete scenario of initial conditions.

The plan of the paper is as follows. In Section 2 we briefly
recapitulate the relevant features of string gas cosmology pertaining
the dilaton. In Section 3 we consider unstable brane
configurations as candidates of initial state. Section 4
investigates the time evolution of the dilaton -- internal
consistency conditions and the question of whether favorable values
can be reached naturally from the brane initial state.

\section{String gas cosmology and dilaton gravity}

String gas cosmology is usually discussed within the framework of
dilaton gravity, which is the simplest modification of Einstein
gravity that respects the T-duality symmetry. The replacement of the
full string theory by this low-energy effective theory is motivated
by making the cosmologically not unreasonable assumption of slowly
varying fields (\emph{i.e.}, the adiabatic approximation), which allows
one to stay at tree level in $\alpha'$. Furthermore weak string coupling
(\emph{i.e.}, $g_s \ll 1$) is assumed. Let us recapitulate its main features.

Assuming no flux and critical string dimension, the dilaton gravity
action in the string frame reads
 \be \label{eq:Dgrav}
  S_{\rm{string}} = \frac{1}{2\kappa_D^2}\int d^Dx \sqrt{-G} e^{-2\phi}\Big(R +
  4(\nabla\phi)^2\Big)\ ,
 \ee
where $\kappa_D$ is the $D$-dimensional reduced gravitational
constant, $D$ being the spacetime dimension, $G$ is the determinant
of the background spacetime metric, $R$ the usual $D$-dimensional
Ricci scalar, and $\phi$ the dilaton. Assuming furthermore as in FRW
cosmology a homogeneous spacetime, the background fields $G$ and
$\phi$ are at most functions of time, and the background metric can
be written in the familiar form $ds^2 = -dt^2 +
\sum_{i}a_i^2(t)dx_i^2$. Rewritten in terms of the number of e-folds
$N_i = \log a_i(t)$ and the shifted dilaton $\varphi\equiv 2\phi -
\sum_i N_i$, the action (\ref{eq:Dgrav}) takes a form that is
manifestly invariant under the T-duality transformation $N_i
\rightarrow - N_i,\ \varphi \rightarrow \varphi,\ (\phi \rightarrow
\phi - N_i)$. When coupled to the matter action of a gas of free
strings
 \be \label{Smatter}
  S_m = \int dt \sqrt{-G_{00}}\ F(N_i, \beta\sqrt{-G_{00}}),
 \ee
where $F$ is the string gas free energy, and varied with respect to
the fields, the dilaton gravity action yields the evolution
equations of string gas cosmology \cite{Tseytlin:1991xk}, given by
\bea
  \dot{\varphi}^2 - \sum_{i=1}^{D-1}\dot N_i^2  = e^\varphi E \label{eq:eom1} \\
  \ddot N_i - \dot{\varphi}\dot N_i             = \frac{1}{2}e^\varphi P_i \label{eq:eom2} \\
  \ddot{\varphi} - \sum_{i=1}^{D-1}\dot N_i^2   = \frac{1}{2}e^\varphi E\ \label{eq:eom3} \ .
\eea
Here $E = F + TS$ is the total energy of the string gas and $P_i =
-\partial F/\partial N_i$ the pressure in the $i$-th direction
multiplied by the total volume.

The cosmology that emerges from these equations is determined by the
behavior of the string gas energy and pressure as functions of the
scale factor. Due to the different form of the scale factor
dependence of the energy levels of the winding and momentum modes,
the winding modes give a negative contribution to the total
pressure, while the momentum modes give a positive one. Hence Eq.
(\ref{eq:eom2}) implies that winding modes tend to prevent
expansion, whereas the momentum modes induce it. In the assumed
initial state, where the universe is string scale sized and filled
with a dense string gas with a temperature close to the Hagedorn
temperature, the energy is nearly constant, $E \sim T_HS$, and the
numbers of winding and momentum modes are equal, so that the total
pressure vanishes. As a consequence the scale factor remains
constant on average, making this initial phase a semi-stable period.
The dilaton, however, is slowly decreasing, as will be discussed in
more detail in Section 4. In the absence of winding modes, space is
free to expand, in which case the temperature drops, so that the
massive string modes eventually go out of equilibrium and the
universe enters a standard radiation dominated era. During this
period, the radii evolve as $a_i \sim t^{2/D}$ as is usual during
radiation domination, while the original dilaton $\phi = (\varphi
+\sum_i N_i)/2$ approaches a constant.

\section{Unstable branes as the origin of the string gas}

\subsection{Brane decay and closed string emission}

The D-branes of bosonic string theory are unstable. As a sign of
this, the spectrum of open strings on a brane contains a tachyonic
mode. Supersymmetric Type II theories also contain unstable
(non-BPS) D-branes, and pairs of stable (BPS) D-branes of opposite
charge become unstable at subcritical separation. The branes decay
to closed strings which then interact and thermalize under suitable
circumstances. Thus it is natural to consider them as an initial
state for hot string gas. This is of course not enough -- the
question is what can be gained by introducing unstable branes?

An immediate bonus is that they provide at least one possible answer
to some conceptual shortcomings of string gas cosmology. Since we
believe that the Universe originates from a Big Bang, it has a
finite history. While string gas cosmology may provide a finite
history for three large dimensions, the gas itself is implicitly
assumed to have an infinite history. On the other hand, for unstable
branes it is possible to introduce initial conditions at some finite
point in time. These initial conditions fall into three categories:
(i) a scenario, where the unstable brane is first created as a
condensate of incoming closed strings, (ii) a scenario, where the
unstable brane pops out from imaginary time, (iii) a scenario, where
the brane initial state is prepared by a complex time contour at the
initial spacelike singularity.

The consequence of having different possibilities for controllable
initial conditions for hot string gas is that one can ask if they
will leave an imprint -- even better, a signature of observable
interest. But before getting there, one must examine some intrinsic
consistency conditions in order to classify the allowed
possibilities. This is the goal of the present paper.

Let us briefly review some facts relating to the decay of D-branes
in bosonic string theory\footnote{For an extensive review, we refer
the reader to \cite{Sen:2004nf}.}. Assuming weak coupling $g_s\ll
1$, an unstable D-brane acts as a classical time-dependent source
for closed string fields. The final state for a $p$-dimensional
brane is then a coherent state of closed strings
\cite{Lambert:2003zr}, \be\label{eq:state}
 \ket\psi \sim : \exp \left\{ -i \sum_s \int d^{p+1}x~J_s(x) \cdot \phi_s(x) \right\}: \ket 0 \ ,
\ee where $J_s(x)$ are source terms for closed string fields
$\phi_s(x)$, and the expression contains a sum over all possible
fields. For the so called full brane decay with a finite
characteristic time scale \be\label{lifetime}
  \tau \sim -\ln [\sin (\pi \lambda ) ]
\ee with a tunable parameter $0\leq \lambda \leq 1/2 $ controlling
the lifetime, the source $J_s$ in the exponent of eq.
(\ref{eq:state}) becomes (after a Fourier transformation to energy
coordinate) \be
  \tilde{J}_s = \pi T_p\frac{\sin (E_s\ln (\lambda ))}{\sinh (\pi E_s)} \ ,
\ee where $T_p$ is the tension of the $p$-dimensional brane. The
tension is inversely proportional to the closed string coupling
constant, so that at weak coupling the brane stores a large energy
density.

Consider first non-compact space and branes, so that there are no
winding modes and \be E_s = E_s(N,k_\perp) = \sqrt{ 4l^{-2}_s(N-1) +
\vec k_\perp^2} \ , \ee where $l_s$ is the string length. The total
energy and number of emitted closed strings from the decay of a
D$p$-brane are \cite{Lambert:2003zr} \bea
 \frac{\bar E}{V_p} & = & \MCN_p^2\sum_{N=0}^\infty d(N)\int \frac{d^{25-p}k_\perp}{(2\pi)^{25-p}}~E_s(N,k_\perp)\bar{n}(N,k_\perp) \label{eq:Ebar} \\
 \frac{\bar N}{V_p} & = & \MCN_p^2\sum_{N=0}^\infty d(N)\int \frac{d^{25-p}k_\perp}{(2\pi)^{25-p}}~\bar{n}(N,k_\perp)\label{eq:Nbar} \ ,
\eea where
\be
 \bar{n} (N,k_\perp) = \frac{|\tilde J_s|^2}{2E_s(N,k_\perp)} \ ,
\ee and the overall coefficient abbreviates
$\MCN_p^2=\pi^{11}(2\pi)^{2(6-p)}$. In (\ref{eq:Ebar}) and
(\ref{eq:Nbar}) the sum is over all final closed string states of
symmetric oscillator excitations between left- and right-moving
sectors and $d(N)$ is the density of states at level $N$.

Note that the number distribution $\bar{n}$ deviates from
thermality. However, for large $E_s$,
 \be
 |\tilde J_s|^2\sim e^{-2\pi E_s} \ ,
\ee so that in the decay the production of highest energy states is
close to the thermal distribution at the Hagedorn temperature $T_H=
1/2\pi$. The total number and energy of strings produced in the
decay depends on the contribution from the highest energies, due to
the exponential growth of the density of left-right symmetric closed
string states \be
 d(N)\sim N^{-27/4}e^{4\pi\sqrt N} \ .
\ee
By using $E_s
\sim 2\sqrt N +\frac{\vec k_\perp^2}{2\sqrt N}$ we can evaluate the
large energy behavior of (\ref{eq:Ebar}) and (\ref{eq:Nbar}). After
performing the momentum integrals one finds that the total amount of
energy per unit $p$-volume carried by all the closed string modes
emitted during the rolling of the tachyon \cite{Sen:2002nu} is
infinite for $p\leq 2$. The reason for the divergence is the
breakdown of perturbation theory. Since an unstable D$p$-brane has a
finite energy, the total energy carried by the closed strings cannot
really be infinite, so higher order corrections must cut it off so
that the total emitted energy is finite.

On the other hand, finite result for $p\geq 3$ means that the single
closed string channel does not carry away all of  the initial energy
of the brane. Since multi-string emission channels are suppressed by
powers of the string coupling, the result means that higher
dimensional branes do not decay completely -- the final state
contains a lower dimensional brane. Hence the decay must be
inhomogeneous. The process is not fully understood at the moment
\cite{Sen:1999mh}.

If we have a D$p$-brane with all its tangential and perpendicular
directions compactified on a torus, then it is related to the
D0-brane via T-duality, and hence we expect that similar results
will hold for this system as well. In particular since under a
T-duality transformation momentum along a circle gets mapped to the
winding charge along the dual circle, we expect that (for
low-dimensional branes) all the energy of the D$p$-brane wrapped on
a torus is converted into closed string radiation. In particular,
most of the energy is stored in the highly wound closed string modes
of mass $\sim g_s^{-1}$ \cite{Mukhopadhyay:2002en}.

\subsection{A proposal for the initial state}\label{sec:prop}

For simplicity, we focus on bosonic string theory.  Initially, all
spacelike directions are compactified on the torus $T^{25}$ with
equal string scale radii in all directions. The most natural initial
state would be a space filling unstable D25-brane (or a stack of
them). However, as we have seen before, branes with $p\geq 3$
presumably decay into lower dimensional branes, and the process is
poorly understood at the moment. When the decay process becomes
better known, we expect that space filling branes could serve as an
initial state for string-brane gas cosmology \cite{Alexander:2000xv}.

For a pure string gas without any branes, we are thus directed to
consider lower dimensional branes which decay completely. A simple
initial configuration consists of a D1-brane wrapped in direction
$X^1$, D1-brane wrapped in direction $X^2$ etc. up to D1-brane
wrapped in direction $X^{25}$. The energy gets released
predominantly in the form of wound strings. The net winding number
is zero if there is no background electric field on the branes. For
homogeneity we assumed equal number of D1-branes at every direction,
with all having the same $\lambda$ (decay equally fast).

As mentioned in the introduction, there are different
prescriptions for the brane decay (and the associated rolling of
the tachyonic mode in its effective potential).  One way is to
first form the brane as a time reversed version of the decay.
Another possibility is to adopt a complex time contour integration
prescription for the computation of closed string production in
the decay, essentially corresponding to nucleating the brane from
imaginary time \cite{Lambert:2003zr}. (However, at present time
there exist no proposals for assigning a probability measure for
the nucleation event.) In this case the spacetime is initially
empty, until the brane suddenly appears and decays into closed
strings. The third possibility is to use the time reflection
symmetry of the tachyonic mode and identify the spacetime points
under a (C)PT reflection into a Lorentzian orbifold. The resulting
orbifold has an initial spacelike singularity, where the brane can
be prepared to nucleate by a variant of the complex contour
integration argument of \cite{Lambert:2003zr} (see
\cite{Kawai:2005jx} for more discussion).

The decay can be adjusted by the parameter $0\leq \lambda \leq 1/2$
which controls the lifetime (\ref{lifetime}). However, for a generic
choice of $\lambda$, the lifetime is very short, not too different
from the string scale. Therefore the brane decay is more like an
explosion. Furthermore, the spacetime stress tensor for the outgoing
closed strings (also called tachyon matter \cite{Sen:2002in}) has
been computed and is known to quickly settle to zero total pressure
\cite{Sen:2004nf}, just as in the initial phase of string gas
cosmology. Hence it is also natural to assume that during the brane
decay the volume of the spacelike torus is essentially unchanged.

In the analysis of the decay, one assumes zero coupling for the
closed strings. This is of course an idealization, and the strings
will interact and backreact to the decay. Furthermore, the
interactions will quickly drive the end state of the decay towards
thermality. The thermalization time scale $t_{\rm th}$ is estimated
by \be\label{termalisaatio} t_{\rm th}\simeq (\sigma \bar n
v_\perp)^{-1} \simeq \frac{\sqrt{l_s E_s}}{g^4_s} l_s^{-1}\gg \tau
\simeq l_s^{-1}~, \ee where $\bar n$ is the string density (formally
divergent by (11), but regulated to be $\sim  l^{-3}_s$), $\sigma$
is the  cross section for string interactions, $\sim g^4_s l^2_s$
for an interaction with a two-string final state \cite{Polchinski,Danos:2004jz},
and $v_\perp \sim 1/\sqrt{l_s E_s}$ is the velocity of the slowly moving
heavy strings of mass $E_s$. (An inspection of the distribution of
the produced closed strings indicates that it is not very far from
thermal distribution at Hagedorn temperature, to start with.) This
crude estimate of the string interaction rate suggests that in the
weak coupling limit the thermalization timescale is much longer than
the brane decay timescale. Hence brane decay and the subsequent
thermalization of closed strings can be discussed separately. After
the thermalization, the standard string cosmological evolution as
described in (\ref{eq:eom1})-(\ref{eq:eom3}) presumably takes over.
Obviously we are assuming that the earlier stages, brane decay and
the dynamics of thermalization, have a minimal effect on the initial
thermal state of the string gas. This could be challenged in many
ways. Even so, there is at least one interesting effect to address.

During the decay, the brane sources the low-energy effective fields,
in particular the dilaton time evolves. One might worry that the
dilaton ends up being too large contradicting the initial assumption
of weak string coupling and insignificant backreaction. More
importantly for the string gas, previous studies of the evolution
equations (\ref{eq:eom1})-(\ref{eq:eom3}) have found that there is a
narrow window for preferred initial conditions of the dilaton in
order to avoid too early freeze-out or too many large space
directions in the end \cite{Easther:2004sd,Danos:2004jz}. An interesting
potential application of the added brane decay stage would be to
drive the dilaton into the preferred initial range from generic
initial values.

\section{Dilaton time evolution and the dimensionality of spacetime}

\subsection{Time evolution of the dilaton during brane decay}

So far we have argued that our proposed setup of decaying branes
produces a universe that looks qualitatively like the initial state
of string gas cosmology. In addition, we need to make sure that the
process is internally consistent with respect to the weak coupling
assumption, and that it creates a final state that fulfills the
adiabatic and weak coupling assumptions of dilaton gravity.

During the brane decay, the evolution of the dilaton is governed by
an equation of motion, which is given by the details of the decay
scenario. After the decay, there follows a period of thermalization,
during which the exact evolution of the dilaton is unknown. However,
one can argue that the explicit form of the distribution of the
background strings has a only a small effect on the propagation of
the dilaton. What counts are the ensemble averaged quantities such as the
mean comoving energy of the gas, which does not change during thermalization.
Hence, we are led to assume that the
dilaton should reach values that are consistent with the dilaton
gravity era already at the end of the brane decay process.

Assuming any backreaction that the produced strings might generate
during the decay process is negligible (as argued in the preceding
Section \ref{sec:prop}, this is the case at least for thermal
backreaction), the dilaton equation of motion during brane decay
reads
 \be\label{yhtalo}
  -\partial^2_t \phi = a\Big[ \frac{1}{1+\hat\lambda e^{t}}+
  \frac{1}{1+\hat\lambda e^{-t}}-1\Big]\ ,
 \ee
where $\hat\lambda=\sin\pi\lambda$ and $a$ is a positive constant,
which is related to the initial tension of the unstable brane. From
now on we set $a=1$. Solving Eq. (\ref{yhtalo}) the time dependence
of the dilaton is found to be
 \be \label{phi}
  \phi(t) = Li_2(-\hat\lambda^{-1} e^{t}) - Li_2(-\hat\lambda e^{t})+C_1 t +C_2\ ,
 \ee
where $Li_2(z)$ is the dilogarithm, and $C_1$ and $C_2$ are
constants of integration. In our notation, the branes begin to decay
at $t=0$ and have lifetimes of $\tau = -\log\hat\lambda$, which are
input parameters. The constants $C_1$ and $C_2$ in Eq.~(\ref{phi})
are determined by assigning to $\phi(t)$ and $\dot{\phi}(t)$ some
initial or final values (corresponding to initial values for the
string gas cosmology era).

The dilaton gravity approximation is valid for $\dot\varphi(t)
\gtrsim -1$ \cite{Easther:2004sd} and, in addition, there are some
constraints on the dilaton following from the equations of motion
(\ref{eq:eom1})-(\ref{eq:eom3}). Firstly, since the energy $E$ is
positive, Eq. (\ref{eq:eom1}) implies that the dilaton time
derivative $\dot{\varphi}$ can never reach zero and must be strictly
positive or strictly negative at all times. Usually $\dot{\varphi} <
0$ is assumed, since $\dot{\varphi}> 0$ could spoil the adiabatic
and weak coupling assumptions. Secondly, it follows from Eq.
(\ref{eq:eom3}) that $\ddot{\varphi}> 0$, and thus $\dot{\varphi}$
grows with time and approaches 0 as $t \rightarrow \infty$. Thus it
is relevant to consider final values for the dilaton time derivative
in the range $-1 \lesssim \dot{\varphi}(\tau) < 0$. Note that the
dilaton solution above is for the string frame dilaton $\phi$,
whereas these values are for the shifted dilaton $\varphi = 2\phi -
\sum_i N_i$. Given that the brane decay is very fast, $\tau\simeq
l_s^{-1}$, it is nevertheless reasonable to assume that the radii
stay fixed during the brane decay, in which case these are simply
related as $\dot{\varphi}(\tau) = 2\dot{\phi}(\tau)$. As for the
value of the dilaton itself, both the brane decay process and the
string gas cosmology scenario require it to be negative, in order to
preserve weak coupling.

\begin{figure}[h!]
\centerline{\epsfxsize=10cm \epsfysize=7cm \epsfbox{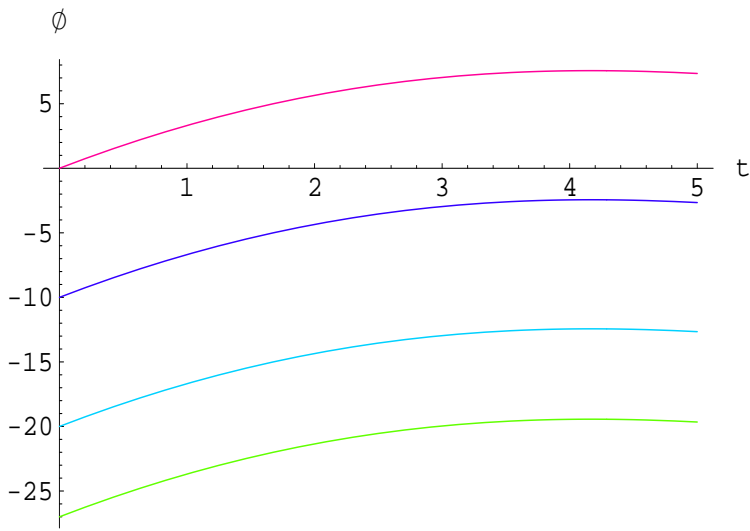}}
\caption[a]{$\phi(t)$ plotted for initial values $e^{\phi(0)} \simeq 1,\
10^{-4},\ 10^{-8}$ and $10^{-12}$, with lifetime $\tau = 5t_s$ and
$\dot{\varphi}(\tau) = -1$.}
\end{figure}

\begin{figure}[h!]
\centerline{\epsfxsize=10cm \epsfysize=7cm \epsfbox{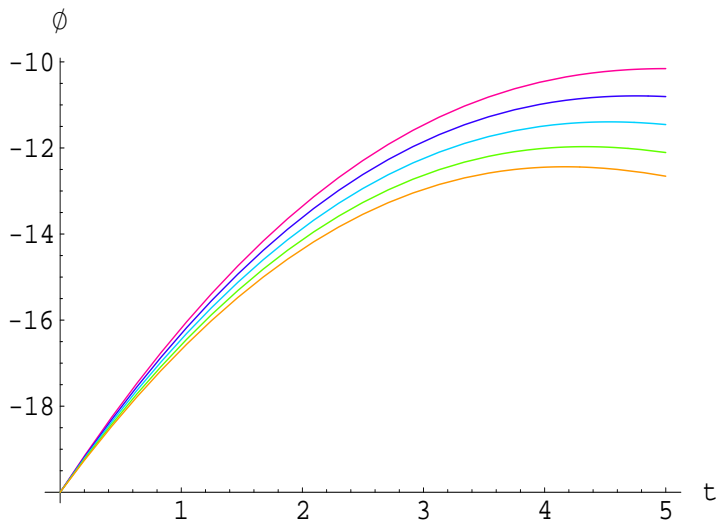}}
\caption[a]{$\phi(t)$plotted for $\dot{\varphi}(\tau) = 0,\ -0.25,\ -0.5,\
-0.75$ and $-1$, where the uppermost curve corresponds to $\dot{\varphi}(\tau) = 0$.
Here we have chosen $\phi(0) = -20$, corresponding to $e^{\phi(0)} = 10^{-8}$, and
$\tau = 5t_s$.}
\end{figure}

\begin{figure}[h!]
\centerline{\epsfxsize=10cm \epsfysize=7cm \epsfbox{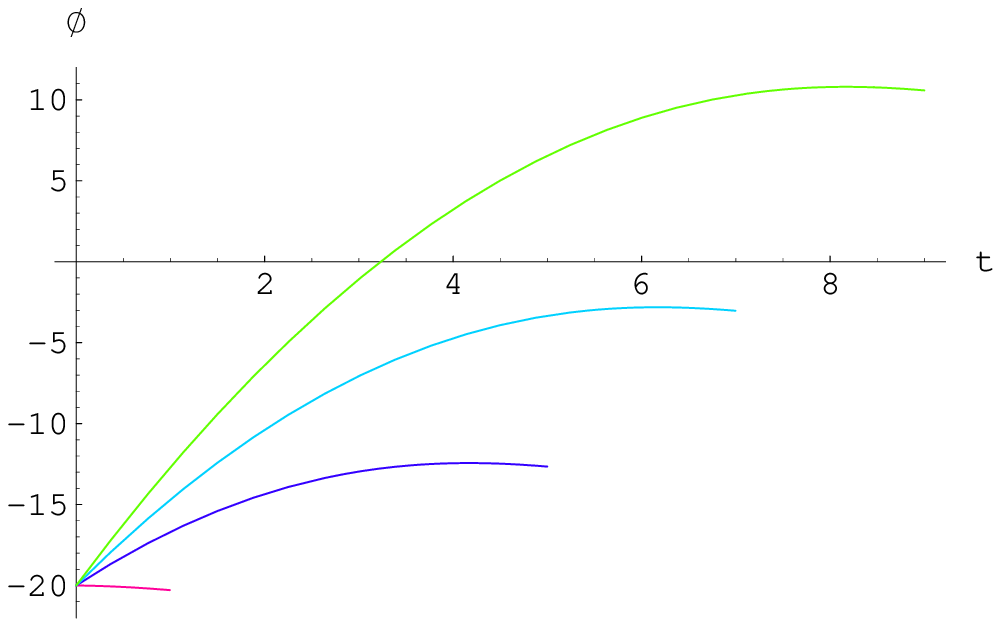}}
\caption[a]{$\phi(t)$ plotted for $\tau = 1,\ 5,\ 7$ and $9$,
with initial value $\phi(0) = -20$ and $\dot{\varphi}(\tau) = -1$. Note
that each curve ends at the value of $t$ corresponding to the decay time $\tau$.}
\end{figure}

In Figs. 1 and 2, we have plotted the dilaton evolution during brane
decay for different initial values and for different final values of
its derivative, respectively. The figures address the consistency of
the weak string coupling. As shown in Fig. 1, starting from large
negative values of the dilaton, the typical decay time (here $5t_s$)
is too short for the dilaton to grow significantly. On the other
hand, Fig. 2 shows that different choices for dilaton time
derivatives at the end of the decay cause no significant change in
the evolution either. Fig. 3 illustrates the dependence of the
dilaton evolution on the lifetime of the decaying branes.

\subsection{Can the dilaton evolution lead to three large dimensions?}

String gas cosmology has been advocated as a mechanism that
dynamically generates a universe with precisely three large spatial
dimensions \cite{Brandenberger:1988aj, Tseytlin:1991xk}. This
argument is based on the observation that the winding modes give a
negative contribution to the string gas pressure, and thus oppose
expansion. In order for a spatial dimension to be able to grow
large, the winding modes wrapped around this dimension must
therefore be annihilated by intersecting with winding modes of
opposite orientation. Since string worldsheets are two-dimensional,
the argument goes, a pair of strings have non-zero probability of
intersecting only in four or less spacetime dimensions, so that at
most three spatial dimensions can grow large. The conclusion of this
qualitative consideration has been confirmed in various studies
\cite{Sakellariadou:1995vk, Alexander:2000xv}, but when cosmological
dynamics, in particular the effect of the dilaton, are taken into
consideration, the simple argument no longer holds.

In a numerical study of the Boltzmann equations governing the string
annihilation in a dilaton gravity background \cite{Easther:2004sd},
it was found that the coupling to the dilaton, which is rolling
towards weak coupling, in general causes the strings to freeze out
too fast for the anisotropic annihilation to take place. In
particular, for initial conditions that admit a large number of
winding modes (\emph{i.e.}, for small initial values of the
dilaton), the strings tend to freeze out so that all dimensions
remain small, whereas for a small initial number of winding modes
(large dilaton) all strings typically annihilate and the whole
compact space grows large. Only for a very narrow range of
intermediate initial conditions is it likely that three dimensions
grow large.

In the previous Section, we studied the evolution of the dilaton
during brane decay, and found that it can indeed lead to values that
are consistent with the dilaton gravity picture. Let us now take one
step further and investigate under what circumstances it might lead
to the values favored for three dimensions to grow large. As argued
above, it is reasonable to assume that the dilaton gravity era and
the study of the Boltzmann equations set in just after the branes
have decayed. In the analysis of \cite{Easther:2004sd}, the
direction of time is chosen so that $\dot{\varphi}<0$, as is done
here. Furthermore, the initial value of the derivative is chosen to
be $\dot{\varphi(\tau)} = -1$, which is just the borderline value
allowed by the dilaton gravity approximation. In this case, there is
a gap of initial values $\Delta\varphi(\tau) \simeq 0.5$, around the
value $\varphi(\tau)\sim-2.5$, for which three dimensions are likely
to become unwrapped \cite{Easther:2004sd}.

\begin{figure}[t]
\centerline{\epsfxsize=10cm \epsfysize=7cm \epsfbox{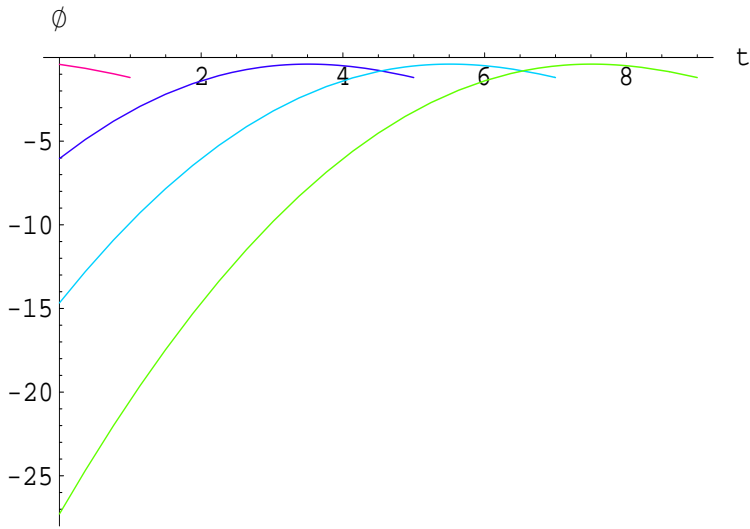}}
\caption[a]{$\phi(t)$ plotted for initial values $\phi(0) = -0.5, -6, -15, -27$,
with $\tau = 1, 5, 7, 9$, correspondingly, and $\dot{\varphi}(\tau) = -1$.}
\end{figure}

In Fig.~4, we have plotted a number of combinations of $\phi(t)$ and
$\tau$, all of which lead the dilaton to values within the favored
range. Thus we see that it is possible to reach the favored range of
initial values for a variety of decay configurations. In particular,
the figure shows that one can start from very weak coupling and
still reach the preferred range for a decay time $\tau \simeq 10t_s$
that is short in comparison to the time of thermalization, which
once again verifies that the decay may be discussed separately from
the other dynamics. Thus we may conclude that in our scenario the
somewhat arbitrary fine-tuning of the dilaton initial value that is
required in the study of \cite{Easther:2004sd} in order to generate
three large dimensions, is given a more physical interpretation in terms
of the lifetimes of the unstable branes.

\section{Discussion}

In this paper we discussed the issue of the origin of thermal
string gas, whose existence is usually assumed in string gas
cosmology. We studied low-dimensional unstable D-branes wrapped on a torus
in bosonic string theory, and found that the string gas is naturally
produced by the decay process of the branes. There are several appealing
features in brane decay that will carry over to string gas cosmology.
For this scenario to hold up, however, it was essential to establish that
the model is internally consistent. We focused on the time evolution
of the dilaton and the string coupling, and found that the evolution is
internally consistent and can lead to favorable values of the dilaton for
three dimensions to grow large. Brane decay thus provides a natural
initial condition for the string gas, and in particular, the initial
values of the dilaton can now be given a physical interpretation in terms
of the lifetimes of the decaying branes.

While such features certainly seem promising, it should be borne
in mind that there are a number of other open issues that must be resolved in order
to fully understand the remnant of brane decay. We initially assume a weak string coupling
so that any backreaction can be neglected and focus only on low-dimensional branes
that are wrapped on a torus with all its radii equal. For homogeneity, we made the assumption
that there is an equal number of branes at every direction of the torus, each decaying
equally fast. Lifetime being of order string time, we furthermore assumed that the
radii stay fixed during the decay process. We then found that during the decay, the dilaton
grows and the interactions with emitted closed strings should be taken into account. By
estimating the order of magnitude of the string scattering rate we are led to
argue that the thermalization timescale is much longer than the decay time, and hence
treating the brane decay and string dynamics separately is justified. Obviously
these assumptions can be challenged in many ways.

An interesting aspect of our proposal is the origin of the entropy
of the string gas. The unstable brane is a coherent state
(\ref{eq:state}), which is a pure state. (More precisely, for the
collection of 25 D1-branes each wound around a different spacelike
circle, the state is a superposition of the coherent states of each
brane, but still a pure state.) On the other hand the state can be
expanded in the basis of the closed string modes, $\ket \psi  =
\sum_s A_s \ket \psi_s$, where the coefficients $A_s$ give
probability amplitudes for decaying into a closed string mode
$\psi_s$. The density matrix $\rho_0$ of the state thus has the
expansion $\rho_0 = \sum_{rs} A_rA^*_s |\psi_r\rangle \langle \psi_s
|$. On the other hand, the thermal gas of strings in the end is
described by the usual canonical ensemble density matrix, a mixed
state with a diagonal expansion in the above basis.

A natural suggestion for making contact with the mixed state of the
thermal gas is to adopt a variant of the coarse-graining proposal of
\cite{rpa}, which considered gravitons produced as a squeezed (pure)
state out of initial vacuum. Applying a random phase approximation
eliminates the off-diagonal elements of the density matrix, leaving
a diagonal mixed ensemble with the entropy depending on the average
occupation numbers per mode according to the familiar formula, and
reducing to the thermal entropy as the occupation number spectrum
thermalizes.

In our context, we correspondingly assume that the phases of the
off-diagonal elements are randomly distributed and averaged out.
This leaves only the diagonal elements, and the density matrix will
reduce to that of a thermal canonical ensemble as the string gas
thermalizes. We leave this and other straightforward extensions for
future work.


\section*{Acknowledgements}

N.J. has been in part supported by the Magnus Ehrnrooth foundation.
K.E. is partially supported by the Academy of Finland grant 114419.
L.M. has been supported by the Vilho, Yrj\"o and Kalle V\"ais\"al\"a
Foundation and gratefully acknowledges a grant by the Mikael
Bj\"ornberg Memorial Foundation. This work was also partially
supported by the EU 6th Framework Marie Curie Research and Training
network ``UniverseNet'' (MRTN-CT-2006-035863).


\end{document}